\documentclass[journal]{IEEEtran}
\usepackage[left=0.75in, right=0.75in, top=1in, bottom=0.9in]{geometry}
\usepackage{mathrsfs,amsthm}
\usepackage{cite}
\makeatletter
\def\ps@headings{%
\def\@oddhead{\mbox{}\scriptsize\rightmark \hfil \thepage}%
\def\@evenhead{\scriptsize\thepage \hfil \leftmark\mbox{}}%
\def\@oddfoot{}%
\def\@evenfoot{}}
\makeatother 
\usepackage{graphicx,amsmath,amssymb,stfloats,subfigure,multicol}%amsmath
\usepackage{epsfig,epsf,psfrag,amssymb,amsfonts,latexsym,graphicx,mathrsfs,subfigure}
\usepackage{bbm}
\usepackage{chngpage}
\usepackage[dvips]{color}
\usepackage{textcomp}
\usepackage{hyperref}
\usepackage{url}
\usepackage[hyphenbreaks]{breakurl}
\usepackage[normalem]{ulem}
\usepackage{algpseudocode}
\newsavebox{\ieeealgbox}

\usepackage{array}
\newtheorem{theorem}{Theorem}

 \def\old#1{}    % Please don't remove this... This command includes the text to be deleted.

\usepackage{bm}

\def\nn{\nonumber}
\def\beq{\begin{equation}}
\def\eeq{\end{equation}}
\def\bea{\begin{eqnarray}}
\def\eea{\end{eqnarray}}
\def\ba{\begin{array}}
\def\ea{\end{array}}

\def\bitem{\begin{itemize}}
\def\eitem{\end{itemize}}
\def\ben{\begin{enumerate}}
\def\een{\end{enumerate}}

\def\eg{{\it e.g., \/}}
\def\etal{{\it et al. \/}}

\newcommand{\mbbE}{\mathbb{E}}

\newcommand{\mbbR}{\mathbb{R}}

\newcommand{\Fmsc}{\mathscr{F}}

\newcommand{\Smsc}{\mathscr{S}}

\def\mubf{\hbox{\boldmath$\mu$\unboldmath}}

\def\omegabf{\hbox{\boldmath$\omega$\unboldmath}}

\def\dbf{{\bm d}}

\def\Dbf{{\bm D}}

\begin{document}

\title{Competitive DER Aggregation for Participation in Wholesale Markets}
% \author{Cong Chen,~\IEEEmembership{Student Member,~IEEE,}
%         Ahmed S. Alahmed,~\IEEEmembership{Student Member,~IEEE,}
%         Timothy Mount,
%         and Lang~Tong,~\IEEEmembership{Fellow,~IEEE,}
\author{Cong Chen,
        Ahmed S. Alahmed,
        Timothy D. Mount,
        and Lang~Tong
		% <-this % stops a space
\thanks{\scriptsize
Cong Chen, Ahmed S. Alahmed and Lang Tong (\{cc2662, asa278, lt35\}@cornell.edu) are with the School of Electrical and Computer Engineering, Cornell University,  USA. Timothy D. Mount (\url{tdm2@cornell.edu}) is with the the Dyson School of Applied Economics and Management, Cornell University,  USA.}
}

\maketitle

\begin{abstract}
The problem of the large-scale aggregation of the behind-the-meter demand and generation resources by a distributed-energy-resource aggregator (DERA) is considered. As a profit-seeking wholesale market participant, a DERA maximizes its profit while providing competitive services to its customers with higher consumer/prosumer surpluses than those offered by the distribution utilities or community choice aggregators. A constrained profit maximization program for aggregating behind-the-meter generation and consumption resources is formulated, from which payment functions for the behind-the-meter consumptions and generations are derived. Also obtained are DERA’s bid and offer curves for its participation in the wholesale energy market and the optimal schedule of behind-the-meter resources.  It is shown that the proposed DERA's aggregation model can achieve market efficiency equivalent to that when its customers participate individually directly in the wholesale market. \end{abstract}

\subsubsection*{Keywords:}

distributed energy resources and aggregation, behind-the-meter distributed generation, demand-side management, net energy metering, competitive wholesale market.

\section{Introduction}\label{sec:Intro}
The landmark ruling of the Federal Energy Regulatory Commission (FERC) order 2222 aims to remove barriers to the direct participation of distributed-energy-resource aggregators (DERAs) in the wholesale electricity markets operated by Regional Transmission Organizations and Independent System Operators (RTO/ISO)  \cite{FERC20}.  By leveraging technological advances in metering and telemetry infrastructure, data-driven energy management, and machine learning technologies, a profit-maximizing DERA can aggregate at scale the growing presence of small-sized generation and demand resources in its participation in the wholesale energy, ancillary, and capacity markets.

Since the release of FERC Order 2222, significant concerns have been raised about whether or not DERAs can be profitable from the wholesale market participation \cite{borenstein2021designing}.  In particular, given that behind-the-meter (BTM) prosumers enjoy significant bill savings under the prevailing net energy metering (NEM) tariff,   attracting customers away from their incumbent utilities or community choice aggregators (CCAs) is a significant challenge  \cite{birk2017tso, SCE_FERC2222:21presentation, GundlachDERA:18EnergyLaw}.

This paper focuses on the {\em competitive DER aggregation} of a profit-seeking DERA in the wholesale electricity market.  By competitive DER aggregation, we mean that the DERA must offer its customers higher consumer/prosumer surpluses than those from their incumbent service providers. To our best knowledge, there is no known profit-maximizing DER aggregation mechanisms that are shown to be profitable and competitive with CCAs and utilities’ NEM-based aggregation models.

The challenge of achieving profitable and competitive DER aggregation is twofold.  First, the DERA must realize more surpluses from distributed generation and demand resources than those individual customers can get under the NEM tariff.  Second, a DERA must maximize its profit from its wholesale market participation, gaining surpluses by offering the wholesale market its aggregated distributed generations and flexible demand resources.  To this end, the DERA, as a wholesale market participant, must derive its profit-maximizing offers and bids from its competitive DER aggregation strategy.

\subsection{Related work}

%  The notion of competitive aggregation is hinted in \cite{Gao&Alshehri&Birge:22} where the authors develop a particular aggregation scheme shown to producers receiving higher prosumer surplus than participating directly to the wholesale electricity market. Since individual participation in the wholesale market is unrealistic, our paper focuses on schemes competitive to competing aggregation models, especially those offered by the regulated utility or community choice aggregators.  Indeed, there is no known aggregation mechanism that is shown to be profitable and competitive with utilities' NEM tariffs. 

 There is growing literature on the DER aggregation and wholesale market participation models that broadly fall into two categories.  One  is through a retail market design operated by a distribution system operator (DSO) \cite{manshadi2015hierarchical, haider2021reinventing}, an aggregation/sharing  platform \cite{morstyn2018multiclass, chen2022review}, or a energy coalition such as CCAs \cite{chakraborty2018analysis, han2018incentivizing}. For the most part, these works do not consider {\em active participations} in the wholesale market with the goal of maximizing the profit of the DERA. In particular, in \cite{manshadi2015hierarchical, haider2021reinventing, morstyn2018multiclass}, the DSO or an aggregation platform participates in the wholesale markets with the aggregated net demand and (possibly) net production, treating the wholesale market as a balancing resource. 
 
 Our approach belongs to the second category of DER aggregations, where independent (possibly profit-maximizing) DERAs aggregate both generation and flexible demand resources within the distribution systems, participating directly in the wholesale electricity market.  As a focus of FERC order 2222, this type of DER aggregation has the potential to improve overall system efficiency and reliability. 

Two significant prior works from this type of DER aggregations are Alshehri et al. \cite{Alshehri&etal:20TPS} and Gao et al. \cite{Gao&Alshehri&Birge:22}, where DERA's profit-maximization involving DER in the distribution system and the wholesale market is considered.  In both cases, the authors considered DERA coordinated decentralized aggregation of generation resources, which can be modeled as a Stackelberg game.
 
 While the notion of competitive DER aggregation has not been formally defined, two prior works have developed competitive aggregation solutions \cite{chakraborty2018analysis,Gao&Alshehri&Birge:22}.
In \cite{chakraborty2018analysis}, Chakraborty \etal considers DER  aggregation by a CCA within the distribution system where the authors provide a pricing mechanism that offers its customers superior surplus than that under the NEM tariff of the regulated utilities. 

Most relevant to our work is the DERA's wholesale market participation method developed by Gao, Alsheheri, and Birge \cite{Gao&Alshehri&Birge:22} where the authors consider a profit-seeking DERA aggregating BTM renewable generations offering its aggregated generation resources to the wholesale market.  In particular, the Gao-Alshehri-Birge (GAB) approach guarantees its customers to achieve a surplus equal
to that achievable by their direct participation in the competitive wholesale market.  In other words, the GAB approach is competitive with the most economically efficient participation model.   A significant difference between the approach in \cite{Gao&Alshehri&Birge:22} and this paper is that we formulate a general competitive aggregation problem that includes the regulated utility or CCA.  In achieving DERA's profit maximization objective, our DER aggregation and market participation models are quite different from that in \cite{Gao&Alshehri&Birge:22}.

% % \textcolor{red}{(The classification here is close to GaoAlshehriBirge.  Check the references they have made).}
% While competitive DER aggregation has not been articulated previously, the work of Gao et al. \cite{Gao&Alshehri&Birge:22} is perhaps the first to hint at such a formulation.  In particular, the Gao-Alshehri-Birge (GAB) approach guarantees its customers to achieve a surplus equal to that achievable by their direct participation in the competitive wholesale market.  In other words, the GAB approach is competitive with the most economically efficient participation model.  

The approach proposed in \cite{Gao&Alshehri&Birge:22} follows the earlier work of Alshehri, Ndrio, Bose, and Ba\c{s}ar \cite{Alshehri&etal:20TPS} where a Stackelberg game-theoretic model is used. Both approaches assume that the DERA elicits prosumer participation with a (one-part or two-part) price, and the prosumer responds with its quantity to supply. The DERA optimizes the price offered to its customers based on {\em anticipated} wholesale market price and submits a quantity bid \cite{ Gao&Alshehri&Birge:22} or a price-quantity bid \cite{Alshehri&etal:20TPS} to the wholesale electricity market. However, the DER aggregation model presented here is significantly different from those in \cite{Alshehri&etal:20TPS, Gao&Alshehri&Birge:22}. We assume that the DERA schedules directly customer's consumption while guaranteeing a competitive advantage over alternative aggregation models. In the wholesale market participation, the DERA in our model submits an offer/bid curve (rather than a quantity) bid without having to anticipate the wholesale market clearing price. Only the expected total BTM generation is needed in the DERA's bid in the wholesale market, and the wholesale market efficiency can be achieved. Our market efficiency result is in parallel to the efficiency participation theorem in  \cite{ Gao&Alshehri&Birge:22}, which is not attained by \cite{Alshehri&etal:20TPS}.

For the competitive aggregation, it is necessary to characterize achievable surpluses by alternative aggregation models, especially for the broadly adopted NEM-based pricing schemes.  To this end, we rely on recent work \cite{AlahmedTong2022NEM, alahmed2022integrating} on the optimal prosumer decision under the general NEM-X tariffs.

\subsection{Summary of results and limitations}

To our best knowledge, this paper develops the first profit-maximizing competitive DER aggregation solution for a DERA to participate in the wholesale energy market.  Specifically, we propose a DER aggregation approach based on a constrained optimization that maximizes DERA surplus while providing $\zeta\% $ higher surpluses than that offered by a competing aggregation model, \eg a regulated utility or CCA under the NEM-X tariff.

The proposed DER aggregation solution provides bid/offer curves for the aggregated generation and demand resources without the DERA anticipating the wholesale market clearing prices and available BTM renewable generations.  It is also shown that the derived bid/offer curves for a price-taking DERA in a competitive market achieve the wholesale market efficiency equivalent to that when DERA's (price-taking) customers participate individually in the wholesale market.

Finally, we present a set of numerical results, comparing the DERA surplus of the proposed competitive aggregation solution with those of various alternatives.  We also compare surpluses of prosumers under different retail market participation models, including regulated utility, the CCA, and DERAs.

In presenting the overall aggregation architecture and aggregation solution, we have ignored some practical implementation details. First, we ignore losses in distribution systems, assuming that the customers' net generations/demands are aggregated without loss at the interconnecting point of the transmission system. In practice, some efficiency parameters may be incorporated, similar to the case of storage participation in the wholesale market \cite{CAISOE320_ESDER}.   Second, we assume that the DERA has a contract with the DSO for its usage of DSO's network in the form of fixed costs.  These costs are passed to customers as connection or delivery charges.  Such an assumption is consistent with the existing consumer aggregation model in many US markets \cite{NYISO2020, ISONE21}. Therefore, we do not account for this part of the aggregation cost and DERA's own operation cost.  Third, under FERC order 2222, DSO may have the right to reject cleared bids and offers of a DERA for reliability reasons \cite{FERC20}.  Our model does not capture the impact of such interventions.

\section{Competitive DER aggregation}\label{sec:Retail}
We present DERA models, NEM-based competitive benchmarks, and the optimized DER aggregation here.

\subsection{DERA interaction with DSO \& RTO/ISO}

\begin{figure}[htbp]
    \centering
    \includegraphics[scale=0.37]{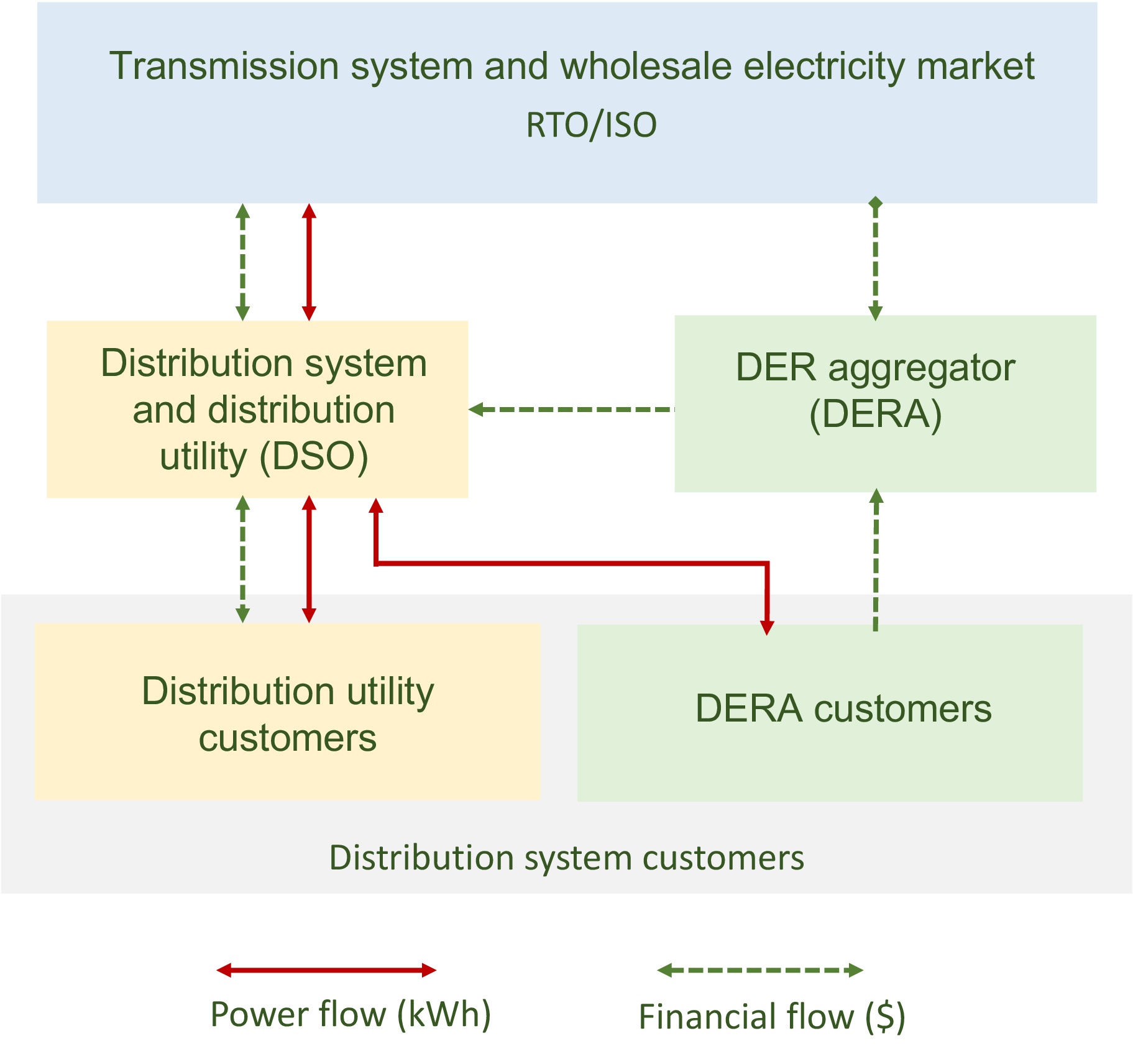}
    \caption{DERA model's physical and financial interactions.}
    \label{fig:DERAmodel}
\end{figure}

We consider a generic interaction model among DERA, DSO, and RTO/ISO shown in Fig. \ref{fig:DERAmodel}, where a DERA uses DSO's physical infrastructure to deliver power to  and from its customers to the wholesale market. It is essential to delineate the financial and physical interactions between DERA and its customers, DERA and RTO/ISO, and DERA and DSO.
\paragraph{DERA and its customer }   We assume a single-bill model for all DERA customers, where each customer is billed for its net consumption by the DERA.   The billing period is consistent with that of the regulated utility for comparison purposes.   The DERA installs an energy management system (EMS) with sensors  and actuators that control major customer consumptions such as heating and air conditioning, water-heater, EV charging, and other controllable energy devices such as battery storage systems.   From its sensors and the reading of its customer's net energy meter, the DERA can account for actual generation and consumption by the customer.

\paragraph{DERA and DSO}  The DERA uses DSO's network and pays DSO for its usage as a fixed cost, which is passed to its customer as a connection charge similar to existing retail tariffs and CCA prices.  The DSO measures the net consumption of the DERA's customers and shares the meter readings with the DERA.  No additional metering is necessary.  The DSO is required to pass the aggregated net consumption to the transmission interconnect, which ensures the strict separation of generations and consumptions from DERA and DSO, thus avoiding double counting.
\paragraph{DERA and RTO/ISO}   The DERA submits its offer/bid curves to the day-ahead or real-time wholesale electricity market (see Sec.~\ref{sec:Whosale}).  The ISO clears the offers/bids and settles payments with the DERA.  In the real-time market, the DERA implements its control based on the cleared market price through its EMSs at the customer sites.

\subsection{NEM benchmarks}

A prosumer in a distribution system can choose to enroll in an NEM X retail program offered by her utility, a CCA, or a DERA providing energy services. A summary of several existing models for the participation of prosumers in utility, CCA and DERA schemes is presented in the Appendix.

In this section, we consider the benchmark performance of a regulated utility offering  the NEM X tariff.  To this end, we leverage the results of \cite{AlahmedTong2022NEM,alahmed2022integrating} and present closed-form characterizations of consumer/prosumer surpluses. 

Consider a prosumer with $K$ energy consuming devices.  Let the $K$-consumption bundle of this prosumer be $\dbf \in \mbbR_+^K$.  We assume that the utility function $U(\cdot)$  is concave, nonnegative, monotonically increasing, continuously differentiable, and additive across the $K$ devices with $U(\dbf)=\sum_{k=1}^K U_k(d_k)$. Let  $V_k(x):=\frac{d}{dx} U_k(x)$ be the marginal utility of consuming $x$ in device $k$.

The prosumer's net consumption is given by
\begin{equation}
   z = \mathbf{1}^\intercal \dbf-g,
\end{equation}
where $g\in [0, \infty)$ is the BTM distributed generation (DG).  The prosumer is a producer if $z<0$ and a consumer when $z\ge 0$.

In evaluating the benchmark prosumer surplus under a regulated utility, we assume that the prosumer maximizes its surplus under the utility's NEM X tariff with parameter $\pi=(\pi^+, \pi^-, \pi^0)$, where $\pi^+$ is the retail (consumption) rate, $\pi^-$ the sell (production) rate, and $\pi^0$ the connection charge.  Under such a tariff, the prosumer's energy bill  $P^\pi(z)$ \cite{AlahmedTong2022NEM} for the net consumption $z$ is given by
 \bea \label{eq:model}
P^{\pi} (z) = \pi^+ [z]^+ - \pi^- [z]^- + \pi^{\mbox{\tiny 0}},
\eea
where $[z]^+=\max\{0,z\}$ and $[z]^-=-\min\{0,z\}$ are the positive and negative parts of $z$, respectively. The prosumer surplus under NEM with parameter $\pi$ is therefore \[S^{\pi}(\dbf):=U(\dbf)-P^{\pi} (z).\]

For an {\em active prosumer} whose consumption is a function of the available DG output $g$, the optimal consumption $\dbf_{\mbox{\tiny NEM-a}}$ and prosumer surplus $S_{\mbox{\tiny NEM-a}}(g)$ can also be obtained by
\beq\begin{array}{l}
\dbf_{\mbox{\tiny NEM-a}} = \arg\max_{ \underline{\dbf} \preceq  \dbf \preceq  \bar{\dbf}} \bigg(U(\dbf)-P^\pi(\mathbf{1}^\intercal \dbf-g)\bigg).\nn
\end{array}
\eeq
The inverse marginal utility (inverse demand curve) for prosumer device $k$ with capacity limits is given by
\beq\label{InverseMU}
f_k(x):=\max \{\underline{d}_k,\min \{V^{-1}_k(x),\bar{d}_k\}\},\eeq
with which the total consumption $d_{\mbox{\tiny NEM-a}} =\mathbf{ 1}^\intercal \dbf_{\mbox{\tiny NEM-a}}$ and the surplus $S_{\mbox{\tiny NEM-a}}(g)$ of  an active prosumer are given  in \cite{AlahmedTong2022NEM} by the following equations.
\begin{align}\label{eq:acitiveSurplus}
    &S_{\mbox{\tiny NEM-a}}(g)=U(\dbf_{\mbox{\tiny NEM-a}})-P^{\pi} (d_{\mbox{\tiny NEM-a}}-g)\\ &=\begin{cases}
\sum_kU_k(d_k^-)-\pi^-(\sum_kd_k^--g)-\pi^0,& g\geq \sum_kd_k^-\nn\\
\sum_kU_k(d_k^+)-\pi^+(\sum_kd_k^+-g)-\pi^0,&  g \le  \sum_kd_k^+\nn\\
\sum_kU_k(d_k^0)-\pi^0, &\text{otherwise}\\
\end{cases}\\
&d_{\mbox{\tiny NEM-a}}= \max \{\sum_{k=1}^Kd_k^+,\min \{g,\sum_{k=1}^Kd_k^-\}\},
% d&_{\mbox{\tiny NEM-a}}&=& \begin{cases} d^-,&g\geq d^-\nn\\
% g,&d^+<g< d^-\nn\\
% d^+, &g\le d^+
% \end{array}
% \eeq
\end{align}
where $d^+_k:=f_k(\pi^+)$,  $d^-_k:=f_k(\pi^-)$, and $d^0_k:=f_k(\mu^*(g))$\footnote{By solving $\sum_{k=1}^Kf_k(\mu)=g$, we have $\mu^*(g)\in [\pi^-,\pi^+]$.} with $f_k(\cdot)$ defined in (\ref{InverseMU}).

We call a prosumer passive if the consumption schedule is independent of the DG output, i.e., all generation is used for bill reductions.  The optimal consumption bundle of such a  {\em passive prosumer} under the NEM X tariff is given by 
\beq\begin{array}{l}
\dbf_{\mbox{\tiny NEM-p}} = \arg\max_{ \underline{\dbf} \preceq  \dbf \preceq  \bar{\dbf}} \bigg(U(\dbf)-\pi^+(\mathbf{1}^\intercal \dbf-g)\bigg).\nn
\end{array}
\eeq
The total consumption $d_{\mbox{\tiny NEM-p}} =\mathbf{ 1}^\intercal \dbf_{\mbox{\tiny NEM-p}}$ and the surplus $S_{\mbox{\tiny NEM-p}}(g)$ of a  {\em passive prosumer} are given by
\begin{align}\label{eq:passiveSurplus}
    &S_{\mbox{\tiny NEM-p}}(g)=U(\dbf_{\mbox{\tiny NEM-p}})-P^{\pi} (d_{\mbox{\tiny NEM-p}}-g)\\ &=\begin{cases}
\sum_kU_k(d_k^+)-\pi^-(\sum_kd_k^+-g)-\pi^0,&g \geq \sum_kd_k^+\nn\\
\sum_kU_k(d_k^+)-\pi^+(\sum_kd_k^+-g)-\pi^0,& g<\sum_kd_k^+ 
\end{cases}\\
&d_{\mbox{\tiny NEM-p}}=  \sum_{k=1}^Kd_k^+.\label{eq:dNEMp}
\end{align}

Both passive and active prosumer classes can be considered in competitive DER aggregation.  In practice, because active prosumer decision requires installing special DG measurement devices and sophisticated control, most prosumers are passive\footnote{Britain establishes a database for passive customers and encourages the participation of passive customers  in the electricity market \cite{Ros18report}. }.

% Let $z^{p}=d^p-g$ be the net consumption pattern of passive producers. 

% \beq\label{passiveSurplus}
% S^{p}=\left\{\begin{matrix}
% U(d^+)-\pi^-(d^+-g)-\pi^0,&if~d^+\le g \\ 
%  U(d^+)-\pi^+(d^+-g)-\pi^0,&if~d^+> g 
% \end{matrix}\right.,
% \eeq

% The optimal  consumption bundle is given by
% \begin{equation}
% \label{eq:OptConsumption}
% \begin{array}{l}
% d^\pi:=\mathbf{1}^\intercal \dbf^*=\left\{\begin{matrix}
%   d^-,& if~  g\geq d^-\\ 
%   g,& if~ d^+ <g< d^- \\   
%   d^+,& if~  g\le d^+  
% \end{matrix}\right.,\end{array}
% \end{equation}

% Additionally a consumer directly joins wholesale electricity market with optimization $\underset{\{P_i^{H}\}}{\rm max}~~ U_i(P^r_i-P^{H}_i)+\pi P^{H}_i-T$ and $d^w_i=max\{\underline{d}_i,min\{\bar{d}_i,U'^{-1}(\pi)\}\}$ will have

% $S_i^{W}(P^r_i)= U_i(d^w_i)+\pi(P^r_i-d^w_i)-T$. 

% When $\underline{d}_i=-\infty, \bar{d}_i=+\infty$, we have $\pi^+ > \pi^- \geq \pi$, thus $U'^{-1}(\pi^+)<U'^{-1}(\pi^-)\le U'^{-1}(\pi) \Rightarrow d^+<d^-\le d^w \Rightarrow U(d^+)<U(d^-)\le U(d^w)$.

% Additionally, a prosumer can chose to participate in a community choice aggregation (CCA) as is illustrated in \textcite{chakraborty2018analysis}. In this case the passive prosumer has surplus 

% $S_i^{pc}=\left\{\begin{matrix}
% U_i(d^+_i)-\pi^-P^{0}_i-T,&if~\sum_i P^p_i\le 0 \\ 
%  U_i(d^+_i)-\pi^+P^{0}_i-T,&if~\sum_i P^p_i> 0 
% \end{matrix}\right.$ 

\subsection{Optimal Competitive DER Aggregation}\label{sec:DERAmodel}

We now consider the optimal DER aggregation of a DERA who maximizes its profit via the wholesale market participation. 

Consider a DERA aggregating $N$ prosumers via centrally scheduling the consumptions of prosumers. In deriving the offers/bids in the wholesale market, the DERA needs to obtain the optimal aggregated production/consumption as a function of the wholesale locational marginal price (LMP) $ \pi_{\mbox{\tiny LMP}}$ at the location of its interconnect with the transmission system.  In particular, the DERA solves for the  consumption bundle of all customers $\Dbf^*$ and their payment functions $\omegabf^*$, defined by
\[
\Dbf^*:=(\dbf^*_n, n=1,\cdots, N),\]
\[
\omegabf^*:=(\omega^*_n, n=1,\cdots, N),
\]
from the following optimization
\bea\label{eq:DERAsurplus_LnGP}
\begin{array}{lrl}
&\underset{\omegabf,\Dbf}{\rm max}&~~ \sum_{n=1}^N (\omega_n-\pi_{\mbox{\tiny LMP}} (\mathbf{1}^\intercal \dbf_n-g_n))\\
&  \mbox{subject to} & \mbox{for all $1 \le n \le N$} \\
& \lambda_n: &{\cal K}_n(g_n) \le U_n(\dbf_n)-\omega_n, \\
% & \eta_i: & \omega^0_i \le P^{NEM}_i,\\
& (\underline{\mubf}_n,\bar{\mubf}_n): &\underline{\dbf}_n \preceq  \dbf_n \preceq  \bar{\dbf}_n,
\end{array}
\eea
where the first constraint, referred to as the {\em ${\cal K}$-competitive constraint}, ensures that the surplus of a prosumer under DERA is $\zeta$-percent markup over the competing prosumer surplus.   For example,  to obtain competitive aggregation over the utility's NEM-based aggregation scheme for {\em passive prosumers},  we have
\beq\begin{array}{l}{\cal K}_n(g_n) =(1+\frac{\zeta}{100}) S_{\mbox{\tiny NEM-p}}(g_n),\nn\end{array}\eeq
which guarantees that the prosumer surplus is at least $\zeta$ percent higher surplus under NEM X. Similarly with (\ref{eq:acitiveSurplus}), competitiveness over {\em active prosumers} can be preserved by ${\cal K}_n(g_n) =(1+\frac{\zeta}{100}) S_{\mbox{\tiny NEM-a}}(g_n)$.

Note that the solution of the above optimization, in general,  implies that the optimal consumption $\dbf_n^*$ is a function of the local DG $g_n$, which would complicate the use of the above optimization as a way to construct offers/bids in the wholesale market, because the offering/bidding process for the wholesale market auction is prior to the realization of these random quantities. Fortunately, such a concern is unwarranted  as the optimal consumption is independent of $g_n$.

\begin{theorem}[{Optimal DERA scheduling and payment}] \label{thm:DERA} 
Given the wholesale market LMP $\pi_{\mbox{\rm\tiny LMP}}$, the optimal consumption bundle $\dbf^*_n=(d_{nk})$ of prosumer $n$, and its payment $\omega_n^*(\dbf_n^*,g_n)$ are given, respectively, by
\beq\label{eq:DERAFixRateGP}
\begin{array}{lrl}
% d^*_{kn}&=& \max \{\underline{d}_{nk},\min \{V_{kn}^{-1}(\pi_{\mbox{\tiny LMP}} ),\bar{d}_{nk}\}\}\\
d^*_{kn}(\pi_{\mbox{\rm\tiny LMP}})&=& f_{nk}(\pi_{\mbox{\rm\tiny LMP}} )\\
\omega^{*}_n(\dbf_n^*,g_n)&= &U_n(\dbf_n^*)-{\cal K}_n(g_n), 
\end{array}
\eeq
where $f_{nk}(\cdot)$ defined in (\ref{InverseMU}) is the inverse demand curve for the $k$-th device of prosumer $n$ .%$\underline{d}_{nk}, \bar{d}_{nk}$ are repectively the lower and upper consumption limits
\end{theorem}
The proof follows directly the KKT conditions of (\ref{eq:DERAsurplus_LnGP}). Note that the optimal consumption for each customer $(d_{nk}^*)$ are only a function of the wholesale market LMP, $\pi_{\mbox{\tiny LMP}}$, independent of the local BTM DG $g_n$. The optimal payment $\omega^*_n$, naturally, depends on both the consumption and the BTM DG $g_n$. The significance of such dependencies is that, once the wholesale market LMP is realized, the scheduled consumption is optimal, unlike the case in \cite{ Alshehri&etal:20TPS, Gao&Alshehri&Birge:22} where the optimally scheduled consumption depends on the forecast of BTM DG $g_n$ and anticipated LMP.

Also significant is that the solution given in Theorem \ref{thm:DERA} provides directly the bid/offer curves that the DERA submits to the wholesale electricity market. Detailed illustrations of the bid/offer curves and  wholesale market efficiency are explained in Sec.~\ref{sec:Whosale}.

% \begin{theorem}[{Optimal DERA scheduling and payment}] \label{thm:DERA} The optimal prosumer scheduling plan, DERA total payment and surplus of DERA from (\ref{eq:DERAsurplus_LnGP}) is $\forall k,n$
% \beq\label{eq:DERAFixRateGP}
% \begin{array}{lrl}
% d^*_{kn}&=& \left\{\begin{matrix}
% \bar{d}_{kn}&if~\pi^w \le V_{kn}(\bar{d}_{kn}) \\ 
% V_{kn}^{-1}(\pi^w),&if~ V_{kn}(\bar{d}_{kn})\le \pi^w \le V_{kn}(\underline{d}_{kn})\\
% \underline{d}_{kn},&if~ \pi^w \geq V_{kn}(\underline{d}_{kn})
% \end{matrix}\right.,\\
% \omega^{*}_n&=& U_n(\dbf_n)-{\cal K}_n ,\\
% S^a&=&\sum_{n=1}^N(\omega^{*}_n-\pi^w (\mathbf{1}^\intercal \dbf^*_n-g_n) -\pi^0).
% \end{array}
% \eeq 
% \underline{If $P^r_i< d_i^+$,}
% \beq\label{eq:DERAFixRateGP2}
% \begin{array}{lrl}
% P^{p*}_i&=& U_i'^{-1}(\pi)-P^{r}_i,\\
% \omega^{*}_i&=&
% U_i(U_i'^{-1}(\pi))-U_i(d_i^+))-\pi^+(P^r_i-d_i^+)\\ 
% \Pi^*&=&\sum_i \Pi_i=\sum_i(\omega^{*}_i-\pi P^{p*}_i ).
% \end{array}
% \eeq 
% \end{theorem}
% {\em Proof:} See Appendix. \hfill $\Box$
% % $P^{p*}_i$ is optimal net consumption of prosumer $i$, and $P^{d*}_i$ is optimal consumption of prosumer $i$ with $P^{d*}_i=U_i'^{-1}(\pi)$.

% \begin{remark}[{Optimal DERA surplus}] \label{thm:OptOmega}DERA surplus $S^a$ solved by (\ref{eq:DERAsurplus_LnGP}) is the maximum over all possible DERA pricing schemes, when DERA maintains ${\cal K}_n $-competitive to certain schemes like NEM X.
% \end{remark}

\section{DERA in the wholesale market}\label{sec:Whosale}
Having obtained the profit-maximizing DER aggregation, we now turn our attention to DERA's participation in the wholesale market.  Here we assume that the DERA is a price-taker in a competitive market. 

\subsection{DERA offer/bid curves}
DERA can participate in the wholesale market as a virtual storage facility \cite{ISONE21}, which allows DERA to inject power into or withdraw power from the wholesale electricity market.  %This section analyzes the optimal bidding strategies in the wholesale energy market as a price taker in a competitive market setting.  

One form of bids by a DERA is the quantity bid, as in \cite{Gao&Alshehri&Birge:22}, where the DERA submits a quantity to buy/sell power to meet the aggregated demand/supply.  Constructing such offers/bids typically requires optimization based on anticipated market clearing price and distributed generation. The market-clearing LMP is used to settle the supply/demand\footnote{LMP is denoted by $\pi$ is this section for simplicity}.  

We focus on the commonly used price-quantity bids, as in \cite{ Alshehri&etal:20TPS}, where the offer/bid curve is used to express the willingness of DERA to buy/sell its aggregated resources.  In a competitive market, such a price-quantity curve is in the form of the marginal cost of production or marginal benefit of consumption.  For the competitive DER aggregation, we develop such offers/bids from the DERA's constrained optimization displayed in (\ref{eq:DERAsurplus_LnGP}).  

As shown in Sec.~\ref{sec:DERAmodel}, the DERA considered in this paper optimizes its DER resources to buy or sell in the wholesale market.  To this end, the DERA submits a bid curve in the market auction to buy or sell its aggregated resources. Such a bid curve---herein referred to as a supply function\footnote{Here we generalize the terminology of supply function to include both purchase and sell of electricity.}, can be constructed as follows.

Let $Q$ be the aggregated quantity to buy (when $Q>0$) or sell (when $Q <0$) for the DERA and $\pi$ be the wholesale market clearing price (LMP).  Let $G=\sum_{n=1}^N g_n$  be the aggregated BTM generation.
It is known that in a competitive market, a price taking DERA participant bids truthfully with its aggregated supply function  $Q=\Fmsc(\pi) $  given by Theorem 1 in (\ref{eq:DERAFixRateGP}),
\beq
\begin{array}{l}\label{Q-P}
 \Fmsc(\pi)= G-\sum_{n,k}f_{nk}(\pi),
 %-\max \{\underline{D},\min \{\sum_{n=1}^N\mathbf{1}^\intercal\dbf^*_{n}(\pi),\bar{D}\}\},
\end{array}
\eeq
% \beq
% \begin{array}{clc}\label{Q-P}
% % s(\pi)&=& \left\{\begin{matrix}
% % R-\bar{D},&if~f(\pi)\geq \bar{D} \\ 
% % R-f(\pi),&if~\underline{D} \le f(\pi) \le \bar{D}\\
% % R-\underline{D},&if~ f(\pi) \le \underline{D}
% % \end{matrix}\right.
% Q(\pi)=G-\max \{\underline{D},\min \{f(\pi),\bar{D}\}\}.,
% \end{array}
% \eeq
% where $\underline{D}=:\sum_{n=1}^N\mathbf{1}^\intercal\underline{\dbf}_n, \bar{D}:=\sum_{n=1}^N\mathbf{1}^\intercal\bar{\dbf}_n$.
where $f_{nk}(\cdot)$ is defined in (\ref{InverseMU}). Note that the inverse of the supply function  $\Fmsc^{-1}(Q)$ defines the offer/bid curves of the DERA.  Note also that 
the supply function depends on the aggregated BTM generation $G$, which is not known to the DERA at the time of market auction.  In practice, $G$ can be approximated using historical data or by $ N \mbbE(g_n)$ by the 
Law of Large Numbers involving $N$ independent prosumers or via the Central Limit Theorem for independent and dependent random variables \cite{Billingsley:95}.

\subsection{Market efficiency with DERA}
We now establish that the DERA's participation in the wholesale market results in the same social welfare as that when all profit-maximizing prosumers participate in the wholesale market individually.  The following theorem is parallel to that in \cite{Gao&Alshehri&Birge:22} under the model that the DERA (and prosumers) submits its bids and offers in the wholesale market auction.
\begin{theorem}[{Market efficiency}] \label{thm:MarketEfficiency} Let $\mbox{\sf SW}_{\mbox{\rm\tiny Direct}}$ and $S_{\mbox{\rm\tiny PRO}}$ respectively be the optimal social welfare and prosumers' surplus when all prosumers directly participate in the wholesale market. Let $\mbox{\sf SW}_{\mbox{\rm\tiny DERA}}$ and $S_{\mbox{\rm\tiny DERA}}$ respectively be the optimal social welfare and DERA's surplus with DERA's participation on behalf of all aggregated prosumers. Then, we have
$$\mbox{\sf SW}_{\mbox{\rm\tiny Direct}}=\mbox{\sf SW}_{\mbox{\rm\tiny DERA}},~ S_{\mbox{\rm\tiny PRO}}=S_{\mbox{\rm\tiny DERA}}.$$
\end{theorem}

The proof of Theorem~\ref{thm:MarketEfficiency} follows by establishing that the wholesale market clearing problem with the proposed DERA offer/bid curves is the same as that when all prosumers participate directly with their individual offer/bid curves.  

Specifically, a price-taking prosumer $n$ who participates directly in the wholesale market constructs her offer/bid curves by solving the following surplus maximization problem with the given $\pi_{\mbox{\tiny LMP}}$:
%Note that, given $\pi_{\mbox{\tiny LMP}}$, $\dbf^{*}_n(\pi_{\mbox{\tiny LMP}})$ in (\ref{eq:DERAFixRateGP}) is the same as the optimal consumption of the price-taking prosumer $n$ when it directly participates in the wholesale market with the following profit maximization problem,
\beq\label{eq:ProfitMaxDirect}\begin{array}{l}
\underset{\dbf_n}{\rm max}~~ U_n(\dbf)-\pi_{\mbox{\tiny LMP}} (\mathbf{1}^\intercal \dbf_n-g_n),\end{array}\eeq
It turns out that the optimal solution of (\ref{eq:ProfitMaxDirect}) is the same as $\dbf^{*}_n(\pi_{\mbox{\tiny LMP}})$  in (\ref{eq:DERAFixRateGP}), which means that the bid/offer curve of prosumer $n$ is  
\beq
\begin{array}{l}\label{Q-P_prosumer}
 \Smsc_n(\pi)= g_n-\sum_{k}f_{nk}(\pi),
 %-\max \{\underline{D},\min \{\sum_{n=1}^N\mathbf{1}^\intercal\dbf^*_{n}(\pi),\bar{D}\}\},
\end{array}
\eeq
where $f_{nk}(\cdot)$ is defined in (\ref{InverseMU}).  From (\ref{Q-P}), we have $ \Fmsc(\pi)=\sum_n  \Smsc_n(\pi)$. So, when $N$ prosumers participate in the wholesale market indirectly through the proposed DERA model, the offer/bid curve to the wholesale market is the same as that when $N$ prosumers do direct participation. That way, the wholesale market clearing problem with the proposed DERA offer/bid curve is the same as that when all prosumers participate directly in the wholesale market. And we have
\beq
\begin{array}{l}
S_{\mbox{\tiny PRO}}=S_{\mbox{\tiny DERA}}=\sum_n(U_n(\dbf^*_n)-\pi_{\mbox{\tiny LMP}} (\mathbf{1}^\intercal \dbf^*_n-g_n)),\nn\end{array}
\eeq
which means that the total surplus of all prosumers directly participating in the wholesale market equals to the surplus gained by the proposed DERA model with prosumers' indirect wholesale market participations. In the latter case, DERA will split the surplus among itself and all aggregated prosumers. Since DERA is profiting from its participation, the individual prosumer receives less payment in the indirect wholesale market participation model through DERA than that from the direct wholesale market participation. This is not unreasonable. After all, individual prosumers cannot participate directly in the wholesale market.

%Therefore, the total cleared quantity and the total prosumer surplus gained when prosumers participating individually are the same as the quantity cleared and surplus gained by DERA when it participates  on behalf of the prosumers.

% This means that the DERA participation in the wholesale market results in overall market efficiency. More precisely, with the proposed DERA model, the social welfare representing the market efficiency is the same as that when all prosumers directly participate in the wholesale market. 

% The proof follows directly from the fact that the wholesale market clearing problem with the proposed DERA offer/bid curve is the same as that when all prosumers participate directly in the wholesale market. From (\ref{eq:ProfitMaxDirect}), the optimal offer/bid curve for a prosumer $n$ directly participating in the wholesale market is 
% \beq
% \begin{array}{l}\label{Q-P_prosumer}
%  \Smsc_n(\pi)= g_n-\sum_{k}f_{nk}(\pi),
%  %-\max \{\underline{D},\min \{\sum_{n=1}^N\mathbf{1}^\intercal\dbf^*_{n}(\pi),\bar{D}\}\},
% \end{array}
% \eeq
% where $f_{nk}(\cdot)$ is defined in (\ref{InverseMU}). From (\ref{Q-P}), we have $ \Fmsc(\pi)=\sum_n  \Smsc_n(\pi)$. So, when $N$ prosumers participate in the wholesale market indirectly through the proposed DERA model, the offer/bid curve to the wholesale market is the same as that when $N$ prosumers do direct participation. Such a result for market efficiency is also parallel to that in \cite{Gao&Alshehri&Birge:22}.

% \input{WholesaleModelMaker_v4}

% \newpage
\section{Case Studies}\label{sec:CaseStudies}
 Six cases were considered in this work, for which the detailed mathematical formulations are explained in the Appendix. In Case 1, all prosumers chose to stay with utility under NEM 2.0 with Ramsey pricing models from \cite{AlahmedTong2022NEM}. In Case 2, all prosumers chose to participate in a CCA with the aggregation model from \cite{chakraborty2018analysis}. In Case 3 and Case 4, all producers respectively chose to be aggregated by DERA with the two-part pricing \cite{Gao&Alshehri&Birge:22} and the one-part pricing \cite{Alshehri&etal:20TPS}. The initial optimal pricing scheme aims at keeping the surplus of producers under DERA competitive with that when the producer directly participates in the wholesale market. Here we revise the DERA pricing model to be competitive with NEM X for the fairness of comparison, and those revised models were shown in the Appendix. The proposed DERA model was simulated in Case 5 and Case 6, where ${\cal K}_n=S^{\mbox{\tiny NEM-p}}_n(g)$ and  ${\cal K}_n=S_n^{\mbox{\tiny CCA-p}}(g)$, respectively\footnote{Mathematical formulation of $S^{\mbox{\tiny CCA-p}}_n(g)$ is shown in (\ref{CCAsurplus}), which represents surplus of {\em passive prosumers} in CCA.}, to simulate DERA competitive to the NEM X and CCA. Normalized surplus of DERA, consumer, producer and utility are analyzed.

\subsection{Parameter settings}

The simulation included $N$ passive prosumer households with homogeneous quadratic utility function $$U_n(x)=\begin{cases}
\alpha x-\frac{\beta}{2}x^2,&0\le x\le \frac{\alpha}{\beta}\\ 
\frac{\alpha^2}{2\beta},&x> \frac{\alpha}{\beta}
\end{cases}, \forall n,$$ 
where $\alpha=\beta=0.24$. Let $\gamma$ percent of the prosumers be producers, and $1-\gamma$ percent of the prosumers be consumers. $\gamma=1$ and $\gamma=0$ represent the extreme cases that all aggregated customers of DERA are DG adopters and all customers are DG non-adopters, respectively. Let $\pi^0=\$0$, $\pi^+=\pi^{-}+\$0.03/kWh$ for the NEM X tariff, and $\pi_{\mbox{\tiny LMP}}=\$0.03/kWh$ for the wholesale market LMP. Assume $V_{kn}^{-1}(\pi), V_{kn}^{-1}(\pi^+), V_{kn}^{-1}(\pi^-)\in[\underline{d}_{kn} , \bar{d}_{kn}], \forall n, k$ for the boundaries constraints of the prosumer $n$ device $k$. Varying DG generation $g_n\in(0,5] kWh$ and $\gamma\in[0,1]$,  we acquire the normalized DERA surplus in Fig.\ref{fig:WeldareDis}.

%, and in our simulation price elasticity $\varepsilon(\pi^{-})=\varepsilon(\pi)=-0.2$ in certain range reasonable rather than extreme case.

% From (\ref{eq:NEM2}), we know
% $\left\{\begin{matrix} &d^-_i=U'^{-1}(\pi^-)=\frac{\alpha-\pi^-}{\beta}=5/6 kWh\\
% &d^+_i=U'^{-1}(\pi^+)=\frac{\alpha-\pi^+}{\beta}=7/12 kWh\\
% &d^w_i=U'^{-1}(\pi)=\frac{\alpha-\pi}{\beta}=7/8kWh\end{matrix}\right.,\forall i$. 

\subsection{Simulation results}
\subsubsection{Surplus of DERA}

\begin{figure}
\centering
%  \hspace{-6em}
 \scalebox{0.38}{\includegraphics{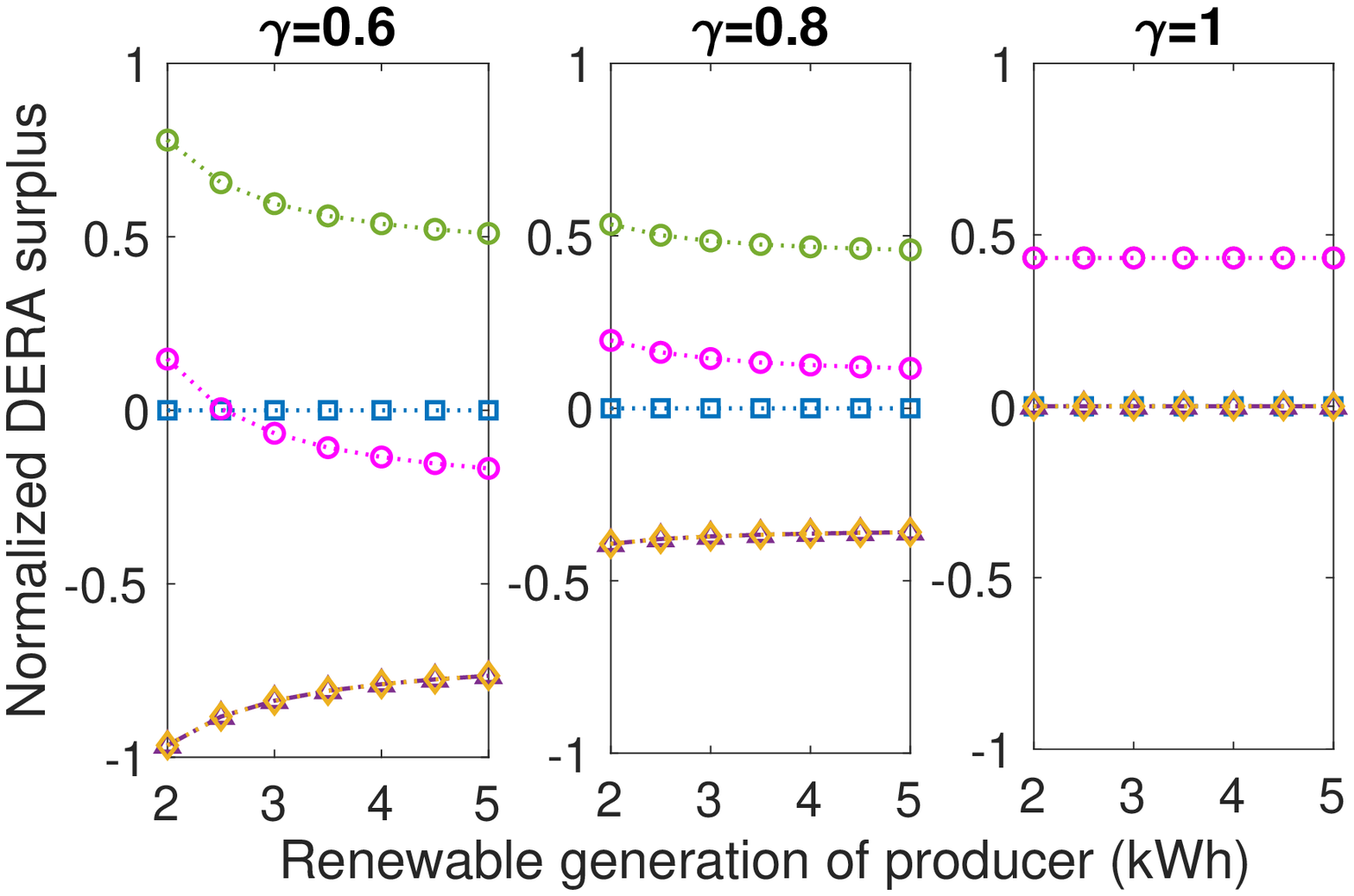}}\\
 \vspace{-2em}
\scalebox{0.38}{\includegraphics{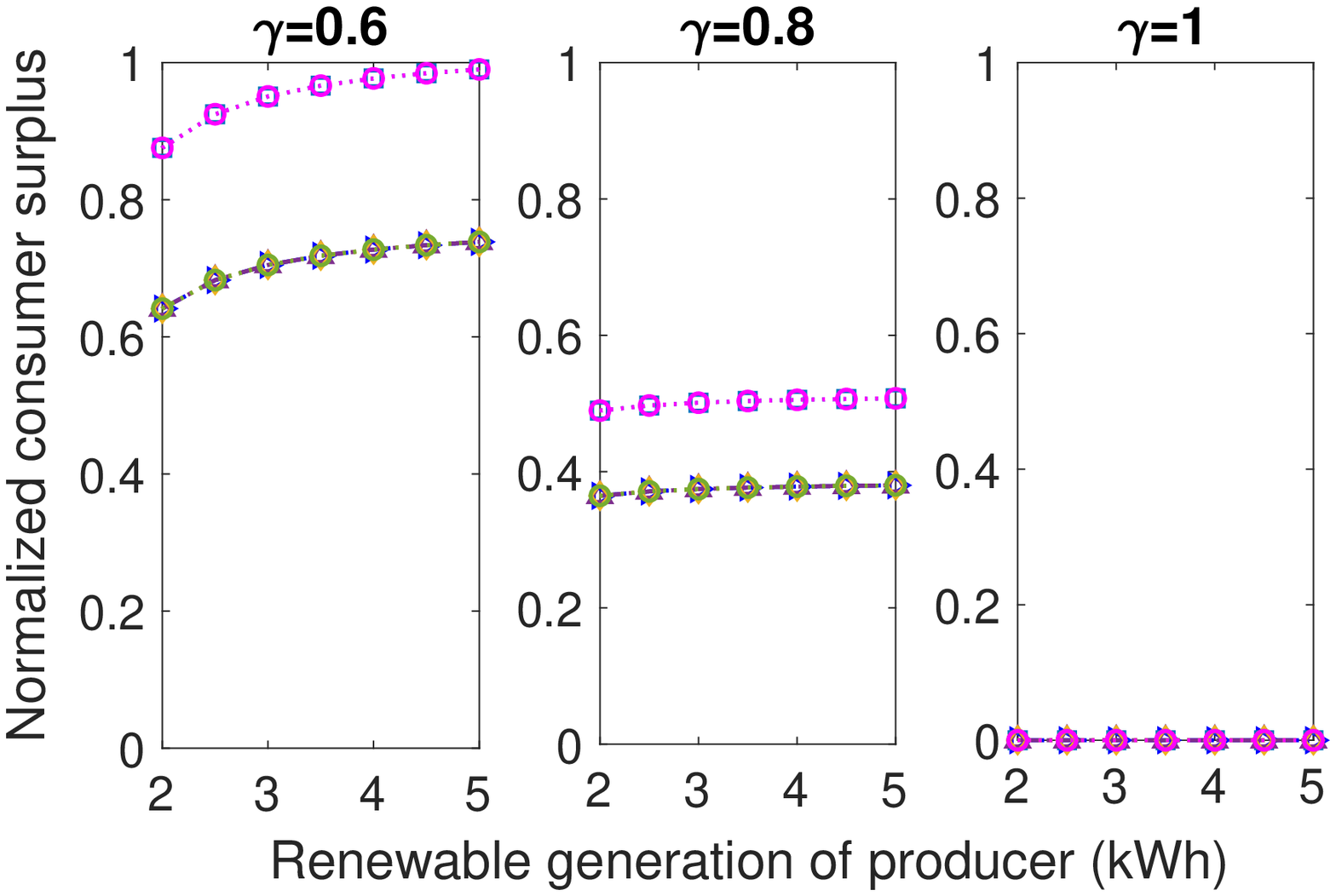}}\\
 \vspace{-2em}
\scalebox{0.38}{\includegraphics{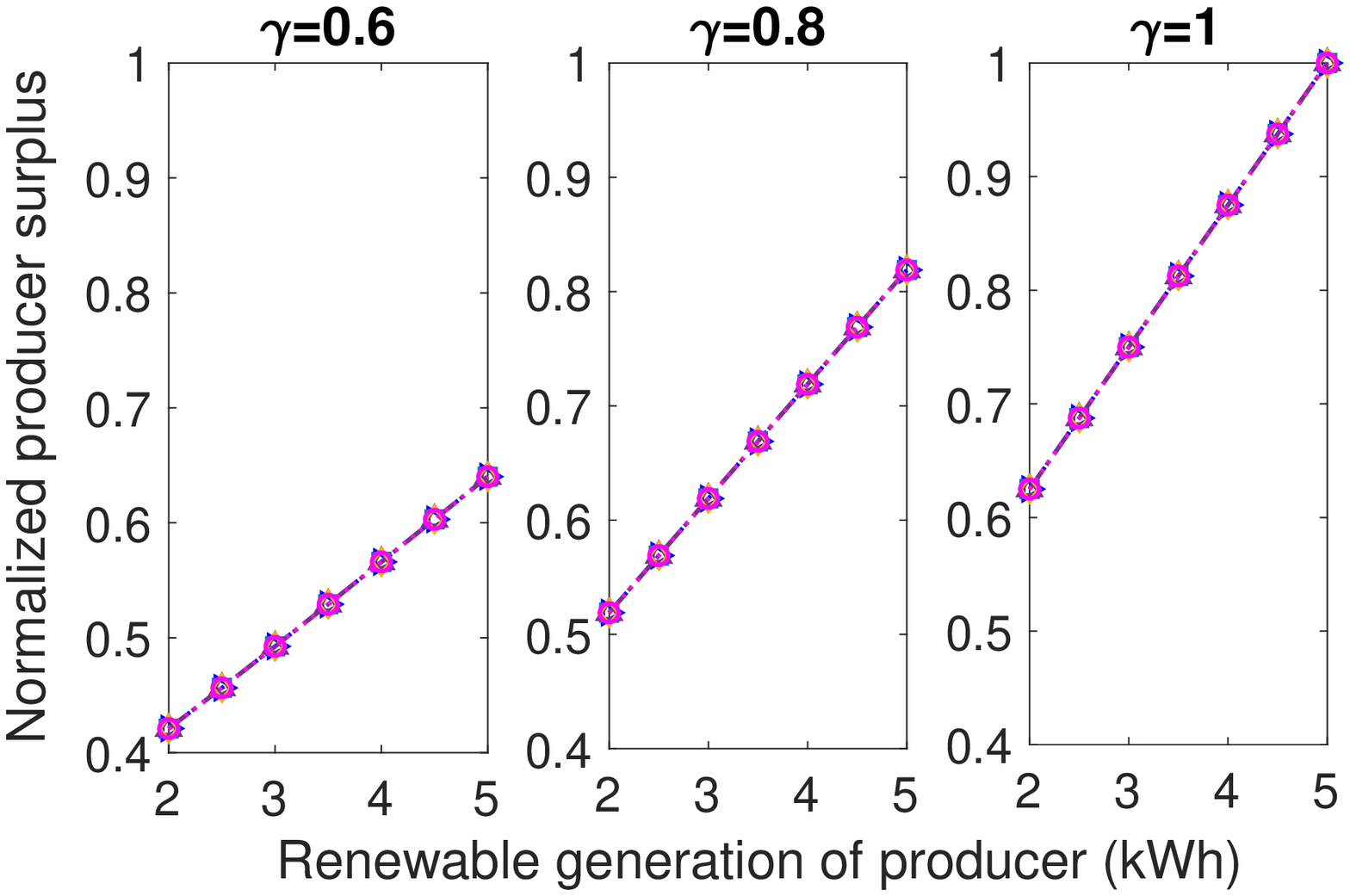}}\\
\vspace{-2em} 
\scalebox{0.38}{\includegraphics{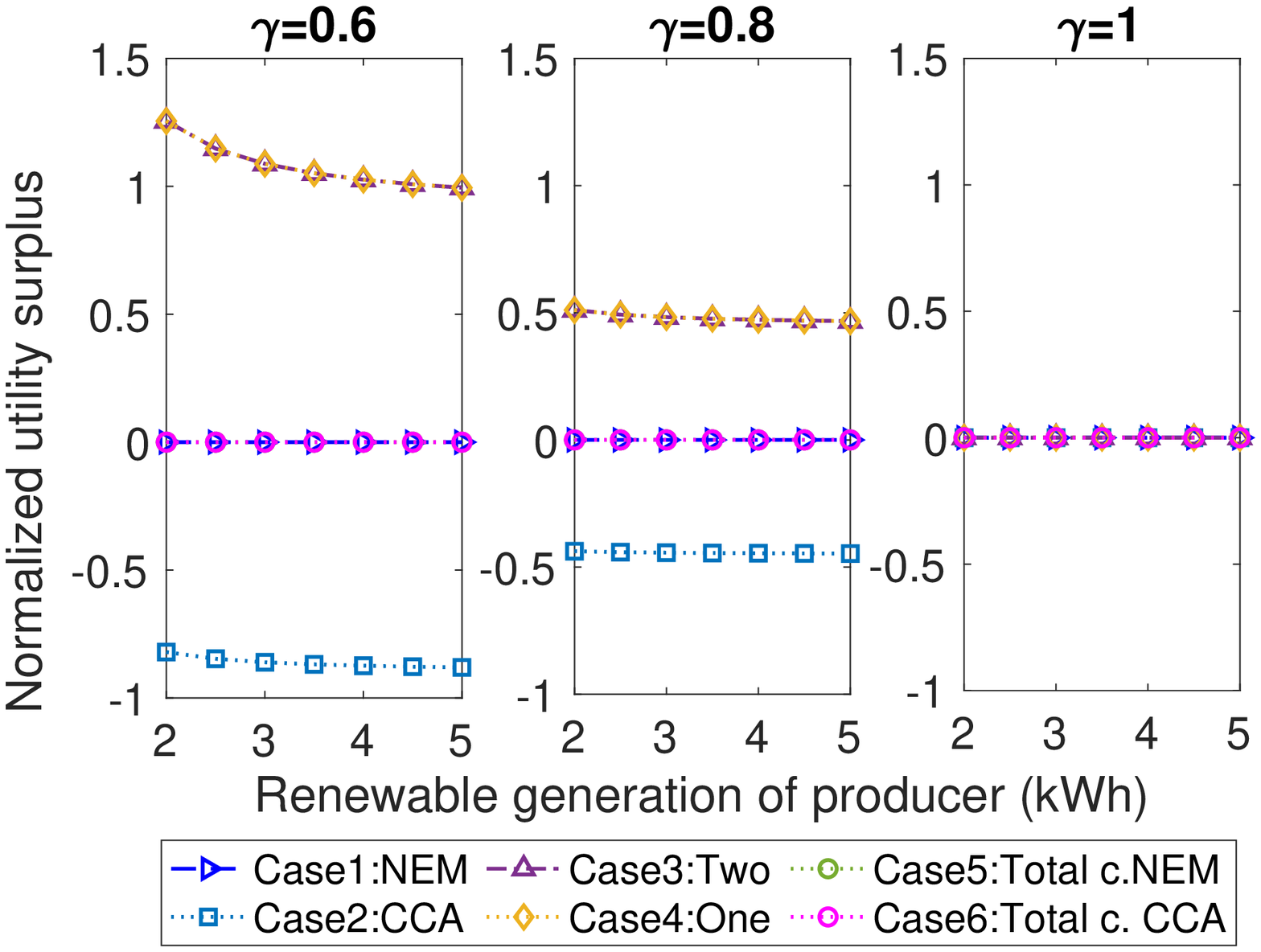}}
 \vspace{-0.5em}
\caption{Welfare distribution.}
\label{fig:WeldareDis}
\end{figure}

It can be observed from the first row of Fig.\ref{fig:WeldareDis} that the proposed DERA model had the maximum DERA surplus over all other cases. When ${\cal K}_n=S^{\mbox{\tiny NEM-p}}_n(g)$ in Case 5, the maximum DERA surplus over all cases was reached. Total surplus of prosumers were larger under CCA than that under NEM X, so when ${\cal K}_n=S_n^{\mbox{\tiny CCA-p}}(g)$ in Case 6, the DERA surplus was smaller than that in Case 5. Although DERA sometimes had negative surplus in Case 5 which is less than the profit neutral CCA in Case 2, surplus of utility in Case 5 was always zero rather than negative in Case 2.

In Case 3 and Case 4, only producers can be aggregated by the proposed DERA model, which was actually a producer aggregator, and it was required that $\pi^+>\pi^-$. So DERA always had negative profit when it was trying to be competitive to NEM X. The more DG generator for the renewable generation, the larger deficiency DERA will have.

\subsubsection{Surplus of producers and consumers}

The second and third rows of Fig.\ref{fig:WeldareDis} show surpluses of consumers and producers, respectively. With the parameter settings, we had negative total net consumption of all $N$ prosumers. It can be observed that consumers in CCA benefited from the aggregation with cheaper price $\pi^-$, and producers in CCA had the same profit as that under NEM X. Thus, among all participation models, consumers had the highest surplus in Case 2 under CCA and in Case 6, which was the proposed method with ${\cal K}_n=S_n^{\mbox{\tiny CCA-p}}(g)$ to be competitive with CCA. All other cases had the same consumer and producer surplus as that under NEM X because ${\cal K}$ competitive constraints were applied with ${\cal K}_n=S_n^{\mbox{\tiny NEM-p}}(g)$ to ensure those models competitive with NEM X for the fairness of comparison.

\subsubsection{Surplus of a regulated utility}

Initially, the utility adopted Ramsey pricing to compute $\pi^+$ and $\pi^-$, where the surplus of utility equaled zero \cite{AlahmedTong2022NEM}. And under the same price $\pi^+$ and $\pi^-$, the utility had a negative surplus with the co-existence of CCA in Case 2, as is shown in the fourth row of Fig.\ref{fig:WeldareDis}. Meanwhile, in Case 3 with the two-part pricing and Case 4 with the one-part pricing, the utility with the same price $\pi^+$ and $\pi^-$ had a positive surplus because all producers chose to be aggregated by DERA leaving all consumers under utility, and we have $\pi^+>\pi$. Therefore, the utility needed to find other ways to maintain profit neutrality in Cases 2, 3, and 4.

In Case 5 and Case 6, the surplus of utility was zero with the proposed DERA model. In these two cases, all prosumers chose to be aggregated by DERA and it's assumed fixed charge of utility equal to the fixed costs for distribution network maintenance, outage services, and so on.%, i.e. $N\pi^0$, from DERA to cover the network operation cost ${\cal C}$, i.e. ${\cal C}-N\pi^0=0$.

\subsection{DERA bids in the wholesale market}

% \subsubsection{Price taker DERA}

\begin{figure}[thb] 
\centering
% \begin{psfrags}
% \scalefig{0.45}\epsfbox{Figs/DERABids3.eps}
% \end{psfrags}
\scalebox{0.55}{\includegraphics{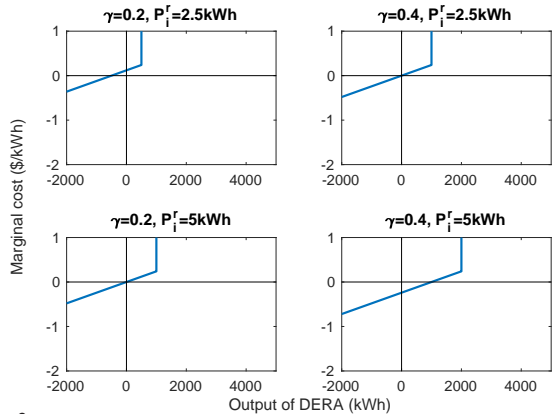}}
\vspace{-1.5em}
\caption{\scriptsize Optimal Bidding curve for price taker DERA.}
\label{fig:DERAbiddingCurve}
\end{figure}

Based on Sec.~\ref{sec:Whosale}, the proposed DERA model had price-quantity bids for a price taker DERA into the wholesale electricity market by submitting the truthful supply function computed from  (\ref{Q-P}), which was 
$$
\Fmsc(\pi)=\begin{cases}
G -\frac{N(\alpha-\pi)}{\beta}, & \pi\le \alpha\\
G, & \pi> \alpha
\end{cases},
$$
where $G=\sum_{n=1}^N g_n$ was the aggregated renewable generation. And the optimal DERA bidding curve\footnote{The DERA bidding curve is the inverse of the supply function.} when $N=1000$ was shown in Fig.\ref{fig:DERAbiddingCurve}. In a competitive wholesale market, DERA revealed those price-quantity curves truthfully to ISO. The slope was determined by parameters of the prosumer utility function $\beta$ and the number of aggregated prosumers $N$. The intersection of the bidding curve with the y-axis was $(0, \alpha-\frac{\beta G}{N})$.
% \scriptsize{
% }

For all price-quantity bids with different parameter settings for $\gamma$ and $g_n$, this DERA can generate to the maximum of $G$ kWh. When $\alpha-\frac{\beta G}{N}<0$, the intersection of the DERA bidding curve with the y-axis is negative. This means that the DERA tended to generate with even a negative price, reflecting redundant renewable energy to be injected to the power network. When $\alpha-\frac{\beta G}{N}>0$, the intersection of the DERA bidding curve with the y-axis is positive. This means DERA tended to withdraw energy from the power network because of the lack of DER generation internally.

\section{Conclusions}\label{sec:Conclusion}
This paper considers the {\em competitive DER aggregation} of a profit-seeking DERA in the wholesale electricity market. As a wholesale market participant, a DERA can both inject and withdraw power from the wholesale market. It is shown that the proposed DERA model maximizes its profit while providing competitive services to its customers with higher surpluses than those offered by the distribution utilities and community choice aggregation.We also establish that the resulting social welfare from DERA's participation on behalf of its prosumers is the same as that gained by the direct participation of price-taking prosumers, making the proposed DERA aggregation model optimal in achieving wholesale market efficiency. Besides, we derive the optimal price-quantity bid of the price-taking DERA in the wholesale market.% which only depends on the aggregated renewable generation and the aggregated optimal prosumer consumption. All the above dependencies ensure that, once the wholesale market LMP is realized, the scheduled consumption is optimal, which is significant for the current wholesale market with ex-ante LMP. And DERA doesn't need accurate price forecast to establish optimal scheduling and pricing plans over the aggregated prosumers.

An open issue of the proposed aggregation solution is that the payment functions for prosumers are nonlinear and non-uniform.  Although each customer is guaranteed to be better off than the competing scheme, two customers producing the same amount may be paid and compensated differently.  In other words, the total charge/credits depend not only on the quantity but also on the flexibilities of the demand and constraints imposed by the prosumer.  Note that a profit-seeking DERA participating in the wholesale electricity market is not subject to the same regulation as a regulated utility.  Such non-uniform pricing is acceptable and has also been proposed in the form of discriminative fixed charges in \cite{Gao&Alshehri&Birge:22}.

% if added before the last page, this command can help balancing columns
%\addtolength{\textheight}{-.2cm} 

%Bibliography 
% \bibliographystyle{apalike}
% \bibliography{sample}

% \printbibliography
{
\bibliographystyle{IEEEtran}
\bibliography{BIB}
}
% \newpage
% \[
% \]
% \newpage

\section{Appendix}\label{sec:Appendix}
\subsection{Participation model of prosumers}\label{sec:Benchmark}
A prosumer participating in the retail market can choose to enroll with the NEM X retail program offered by the utility or enroll with the DERA participation scheme. In this context, a summary of several existing models for the participation of {\em passive prosumers} in the regulated utility, the CCA and DERA schemes are presented \footnote{For simplicity, $U_{i}(d^{\mbox{\tiny NEM-p}}_i)$ in this section represents the total utility over all devices for prosumer $i$, i.e. $\sum_kU_{ik}(d^+_{ik})$.}. %We also establish  model to analyze the welfare distribution among all units

\subsubsection{NEM X}

Consider $N$ prosumers under NEM X. Let $\gamma$ be the fraction of producers indexed by $i$ with $d_i^+-g_i\le 0, \forall i\in\{1,...,N\gamma\}$, and $1-\gamma$ be the fraction of consumers indexed by $j$ with $d^{\mbox{\tiny NEM-p}}_j-g_j\geq 0,\forall j\in\{N\gamma+1,...,N\}$. The surplus of the utility company is
\begin{align}\label{UtilitySurplus}
S^u(\pi^+,\pi^-)&=\sum_{i,j}\bigg(\pi^0+(\pi^--\pi_{\mbox{\tiny LMP}})(d^{\mbox{\tiny NEM-p}}_i-g_i)\nn\\
&+(\pi^+-\pi_{\mbox{\tiny LMP}})(d^{\mbox{\tiny NEM-p}}_j-g_j)\bigg)-{\cal C},
\end{align}
where ${\cal C}$ is the network operation cost of the utility. For simplicity, we can assume $N\pi^0={\cal C}$. %\textcolor{red}{or for risk control of utility based on the variable cost.}

From (\ref{eq:passiveSurplus}), the surplus of passive producing and consuming prosumers are $S^{\mbox{\tiny NEM-p}}_i$ and $S^{\mbox{\tiny NEM-p}}_j$, respectively. The regulator of a profit-neutral utility will construct the NEM X tariff by solving a Ramsey pricing problem proposed in \cite{AlahmedTong2022NEM}.
\begin{equation} \label{RamUltiyNEMEx}
\begin{array}{lrl}&\underset{\pi^+,\pi^-}{\rm max} & \sum_iS_i^{\mbox{\tiny NEM-p}}+\sum_jS_j^{\mbox{\tiny NEM-p}}\\ 
& s.t.& S^u(\pi^+,\pi^-)=0,\\
% d_i^+=\frac{\alpha - \pi^+}{\beta},d_i^+=d_j^+\\
% && \pi^+= \pi^-+0.03\\
% &&\pi \le \pi^-
\end{array}\end{equation}
where the regulator maximizes the social welfare of all participants given that the regulated utility achieves revenue adequacy and profit neutrality.% In practice, we usually have $\pi^+\geq \pi^-\geq \pi_{\mbox{\tiny LMP}}$.

\subsubsection{One-part pricing over producers}

The optimal DERA one-part pricing scheme is proposed in \cite{Alshehri&etal:20TPS}, where producer $i$ can choose to sell energy to DERA with price $\omega^1_i$. The initial optimal pricing scheme aims at keeping the surplus of producers under DERA competitive with that when the producer directly participates in the wholesale market. Here we revise the DERA one-part pricing model to be {\em $\cal K$-competitive} for the fairness of comparison. In this case, the surplus of a passive producer $i$ is given by
$S^{po}_i(g_i)=\left\{\begin{matrix}U_{i}(d^{\mbox{\tiny NEM-p}}_i)- \omega^1_i z^{\mbox{\tiny NEM-p}}_i& z^{\mbox{\tiny NEM-p}}_i<0\\ U_{i}(d^{\mbox{\tiny NEM-p}}_i)-\pi^+z^{\mbox{\tiny NEM-p}}_i,&z^{\mbox{\tiny NEM-p}}_i\geq 0\\ \end{matrix}\right.$,
where $z^{\mbox{\tiny NEM-p}}_n=d^{\mbox{\tiny NEM-p}}_n-g_n$.
% \begin{equation} \begin{array}{lrl}&& \pi_i(P^r_i)=\left\{\begin{matrix} p_j[P^r_i-d^+_i]^++U_i(d^+_i)&if~ P^r_i-d^+_i>0\\ \pi^+( P^r_i-d^+_i)+U_i(d^+_i),&if~P^r_i-d^+_i\le 0\\ \end{matrix}\right.\\ & s.t. &0\le x_j\le P^r_i\\&  & 0\le d_j\le Z-P^r_i+x_j (\Rightarrow 0\le C_j-x_j+d_j\le Z,x_jd_j=0)\\
% &&U_i(d^+_i) +\pi^-(P^r_i-d^+_i) \le p_j[P^r_i-d^+_i]^++U_i(d^+_i)
% \end{array} \end{equation} 

The profit maximization of the DERA to get price $\omegabf^1$ is given by
\begin{equation} \label{oneDERA}
\begin{array}{lrl}&\underset{\omegabf^1}{\rm max} & \sum_j (\pi_{\mbox{\tiny LMP}}-\omega_i^1)[g_i-d^{\mbox{\tiny NEM-p}}_i]^+\\
% &&=\sum_{j\in \{j:\pi^+\geq \frac{\beta_j}{P^r_i}\}}(P_j+(\lambda-p_j)(P^r_i-\frac{\beta_j}{\pi^+}))\\ 
& s.t. &{\cal K}_i\le  \omega^1_i[g_i-d^{\mbox{\tiny NEM-p}}_i]^++U_i(d^{\mbox{\tiny NEM-p}}_i) .
% &&0 \le p_j\le \pi_{\mbox{\tiny LMP}}.
\end{array}\end{equation}

To make the DERA competitive to the NEM X retail program with the surplus of passive conputed in (\ref{eq:passiveSurplus}), we have ${\cal K}_i=S^{\mbox{\tiny NEM-p}}_i$. So the optimal profit of DERA computed from (\ref{oneDERA}) is given by
\begin{equation} \begin{array}{lrl} S^{ao}=\sum_{i}(\pi_{\mbox{\tiny LMP}}-\pi^-)[g_i-d^{\mbox{\tiny NEM-p}}_i]^+,\end{array}\end{equation}
which is independent of $\omega^1_i$. If all producers choose to sell energy to the DERA, the surplus of utility becomes
\beq 
\begin{array}{lrl}\label{UtilitySurplusOne}
S^{uo}=\sum_j(\pi^+-\pi_{\mbox{\tiny LMP}})(d^{\mbox{\tiny NEM-p}}_i-g_j),\end{array}\eeq
which is non-negative as we usually have $\pi^+\geq \pi_{\mbox{\tiny LMP}}$.

\subsubsection{Two-part pricing over producers}

The optimal DERA two-part pricing scheme $(\omega^1,\omega^2_i)$ is proposed in \cite{Gao&Alshehri&Birge:22}.  Similarly, here we revise it to be {\em $\cal K$-competitive} for the fairness of comparison. In this case, with passive producer net consumption $z^{\mbox{\tiny NEM-p}}_i=d^{\mbox{\tiny NEM-p}}-g_i$ in (\ref{eq:dNEMp}), the surplus of a passive producer is given by

$S^{pt}_i(g_i)=\begin{cases} U_i(d^{\mbox{\tiny NEM-p}}_i)-\omega^1z^{\mbox{\tiny NEM-p}}_i-\omega^2_i,&z^{\mbox{\tiny NEM-p}}_i<0\\ U_i(d^{\mbox{\tiny NEM-p}}_i)-\pi^+z^{\mbox{\tiny NEM-p}}_i,&z^{\mbox{\tiny NEM-p}}_i\geq 0\\ \end{cases}.
$\\
% \begin{equation} \begin{array}{lrl}&& \pi_i(P^r_i)=\left\{\begin{matrix} p_j[P^r_i-d^+_i]^+-P_j+U_i(d^+_i)&if~ P^r_i-d^+_i>0\\ \pi^+( P^r_i-d^+_i)+U_i(d^+_i),&if~P^r_i-d^+_i\le 0\\ \end{matrix}\right.\\ & s.t. &0\le x_j\le P^r_i\\&  & 0\le d_j\le Z-P^r_i+x_j (\Rightarrow 0\le C_j-x_j+d_j\le Z,x_jd_j=0)\\
% &&U_i(d^+_i) +\pi^-(P^r_i-d^+_i) \le p_j[P^r_i-d^+_i]^+-P_j+U_i(d^+_i)
% \end{array} \end{equation} 

And the profit maximization of the DERA is 
\begin{equation}\label{twoDERA} \begin{array}{lrl}&\underset{\omega^1,\omegabf^2 }{\rm max} & \sum_i (\omega^2_i \mathbf{1}\{z^{\mbox{\tiny NEM-p}}_i<0\}-(\pi_{\mbox{\tiny LMP}}-\omega^1)z^{\mbox{\tiny NEM-p}}_i)\\
% &&=\sum_{j\in \{j:\pi^+\geq \frac{\beta_j}{P^r_i}\}}(P_j+(\lambda-p_j)(P^r_i-\frac{\beta_j}{\pi^+}))\\ 
& s.t. &{\cal K}_i\le  U_i(d^{\mbox{\tiny NEM-p}}_i)+\omega^1[z^{\mbox{\tiny NEM-p}}_i]^--\omega^2_i.
% &&0 \le p_j\le \lambda.
\end{array}\end{equation}
% $\Rightarrow P^*_j=  (p_j^*-\pi^-)(P^r_i-\frac{\beta_j}{\pi^+})\mathbf{1}\{P^r_i-\frac{\beta_j}{\pi^+}>0\}$.
Similarly, to make DERA competitive to NEM X, we have ${\cal K}_i=S^{\mbox{\tiny NEM-p}}_i= U_i(d^{\mbox{\tiny NEM-p}}_i) +\pi^-(g_i-d^{\mbox{\tiny NEM-p}}_i)$. In this case, optimal profit of the DERA computed from (\ref{twoDERA}) is the same as that from (\ref{oneDERA}), which is independent of $(\omega^1,\omega^2_i)$. The utility's surplus is the same as (\ref{UtilitySurplusOne}).

\subsubsection{Community choice aggregation}

Based on current market rules, prosumers can participate in a community choice aggregation (CCA) to reduce energy bills or better manage roof-top DG. In \cite{chakraborty2018analysis}, a profit-neutral CCA model is proposed. For a passive prosumer participating in a CCA with net consumption $z^{\mbox{\tiny NEM-p}}_n=d^{\mbox{\tiny NEM-p}}_n-g_n$ in (\ref{eq:dNEMp}), the prosumer surplus is
\beq\label{CCAsurplus}
S_{\mbox{\tiny CCA-p}}(g)=\left\{\begin{matrix}
U_n(d^{\mbox{\tiny NEM-p}}_n)-\pi^-z^{\mbox{\tiny NEM-p}}_n-\pi^0,&z_{\mbox{\tiny CCA}}\le 0 \\ 
 U_n(d^{\mbox{\tiny NEM-p}}_n)-\pi^+z^{\mbox{\tiny NEM-p}}_n-\pi^0,&z_{\mbox{\tiny CCA}}> 0 
\end{matrix}\right.,
\eeq
where $z_{\mbox{\tiny CCA}}=\sum_nz^{\mbox{\tiny NEM-p}}_n$.

If all $N$ prosumers participate in CCA and  ${\cal C}=N\pi^0$, the surplus of the utility is
\beq
\begin{array}{lrl}
S^{uc}=\left\{\begin{matrix}\sum_n(\pi^--\pi_{\mbox{\tiny LMP}})(d^{\mbox{\tiny NEM-p}}_n-g_n),&z_{\mbox{\tiny CCA}}\le 0 \\
\sum_n(\pi^+-\pi_{\mbox{\tiny LMP}})(d^{\mbox{\tiny NEM-p}}_n-g_n),&z_{\mbox{\tiny CCA}}> 0 
\end{matrix}\right..
\end{array}
\eeq
% $P^{CCA}(P^r_i)=\left\{\begin{matrix}\pi^+(d_i-P^r_i),& if \sum_i(d_i-P^r_i)>0\\
% \pi^-(d_i-P^r_i),& if \sum_i(d_i-P^r_i)<0
% \end{matrix}\right.$.

% $S_i^{P}(P^r_i)=\left\{\begin{matrix}
% U_i(d^+_i) +\pi^+(P^r_i-d^+_i)-T,&if~ \sum_i(d_i-P^r_i)>0\\ 
%  U_i(d^+_i)+\pi^-(P^r_i-d^+_i)-T,&if~\sum_i(d_i-P^r_i)<0
% \end{matrix}\right.$ 

% \subsection{Proof of Theorem \ref{thm:DERA} }
% \input{appendixProofTHM1_v5}%proofs
% % \input{appendix_v0}%proofs

% \subsection{Proof of Theorem \ref{thm:MarketEfficiency} }
% \input{appendixProofTHM2_v8}

\end{document}